\pdfoutput=1
\documentclass[aps,pre,twocolumn,showpacs,groupedaddress]{revtex4-1}

\usepackage{amsmath}    
\usepackage{verbatim}   
\usepackage{color}      
\usepackage{subfigure}  
\usepackage{hyperref}   
\usepackage{graphics}
\usepackage{epsfig}
\usepackage[english]{babel}
\usepackage{natbib}
\usepackage{fancybox}

\begin{document}

\title{Locally self-organized quasi-critical percolation in a multiple disease model}

\author{Jeppe Juul}
\author{Kim Sneppen}
\affiliation{Niels Bohr Institute, Blegdamsvej 17, DK-2100 Copenhagen, Denmark}

\date{\today}

\begin{abstract}
Diseases emerge, persist and vanish in an ongoing battle for available hosts. Hosts, on the other hand, defend themselves by developing immunity that limits the ability of pathogens to reinfect them.
We here explore a multi-disease system with emphasis on mutual exclusion. We demonstrate that such a system develops towards a steady state, where the spread of individual diseases self-organizes to a state close to that of critical percolation, without any global control mechanism or separation of time scale.
For a broad range of introduction rates of new diseases, the likelihood of transmitting diseases remains approximately constant. 
\end{abstract}


\maketitle

\section{Introduction}
Many phenomena within materials science, physics, and biology are associated with percolation theory \cite{Sahimi, Stauffer}. In particular, it has been shown that bond percolation is equivalent to the class of susceptible/infectious/recovered (SIR) epidemic models on a network \cite{Grassberger, Miller, NewmanA, NewmanB, Sander, Serrano}. 

In such an SIR model, all nodes start out susceptible to a new disease. If a node is infected, it will try to infect its neighbors for a fixed time $\tau$, after which it recovers and becomes immune to the disease. How widely each disease is spread on the network depends on the probability $p$, with which a node infects each of its neighbors before it recovers. The probability $p$ will, therefore, be directly given by the disease time $\tau$. If $p$ is greater than a critical percolation threshold $p_c$, there is a finite probability that the disease will span the entire infinite network, thus becoming an epidemic.

In nature, percolation phenomena are often found near the critical probability $p_c$ \cite{Warren}. 
A possible explanation of this is the concept of self-organized criticality (SOC), where complex systems drive themselves to critical states without the need for fine-tuning of the parameters \cite{Bak, BakSneppen, Paczuski}. 
Many models for self-organized percolation have been studied \cite{Wilkinson, Alencar, BakForest, Henley, Zapperi}. In these, the self-organization either arises as a result of very different time scales or through dynamics involving a global control mechanism. For instance, a percolation system can self-organize to the critical threshold by dynamically adjusting the probability $p$, such that the percolation cluster keeps growing at a specific rate \cite{Alencar}. However, this requires that all nodes on the network 'know' how fast the cluster is growing globally; a condition that is rarely fulfilled.  

In this paper we study an SIR model for the spread of multiple diseases that compete with each other. When many diseases are present, they may well influence each other by weakening of host immunity, or they may inhibit each other through cross-immunization or by mutual exclusion \cite{Pease, Andreasen, Gog, Kryazhimskiy}. 
Considering mutual exclusion only, we here show that the system self-organizes to a state close to the critical percolation threshold for a wide range of input parameters. 
That is, it exhibits self-organized quasi-criticality without any global control mechanism or separation of time scale.

\section{Model}
\begin{figure}[tb]
	\centering
		\includegraphics[width=\columnwidth]{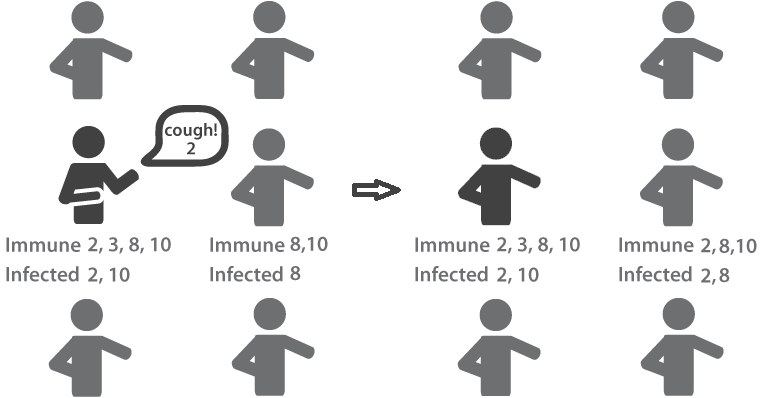}
	\caption{At each time step, a random node 'coughs' on a random neighbor and thereby transmits a random one of its diseases. If the neighbor is not already immune to the disease, it is infected and becomes infectious for a time $\tau$.}
	\label{fig:spreadMechanismLattice}
\end{figure}

In our model, diseases are spread on a 2-d square lattice with periodic boundary conditions and $N = L^2$ sites, each representing a host. At each time step the following actions take place:
\begin{itemize}
	\item With the small probability $\frac \alpha N$, a new disease originates in a random node on the network.
	\item A random node $i$ and one of its four neighbors $j$ are selected. If $i$ carries any disease(s), a random disease is selected. If $j$ is not already infected or immune to this disease, it is transmitted to $j$ (see fig \ref{fig:spreadMechanismLattice}). 
	\item After $\tau$ sweeps over the lattice, $j$ will be cured from the disease.
\end{itemize}
The model uses a framework similar to the one recently developed in \cite{Sneppen}, except that the present model allows each node to be infected by several diseases at the same time. Also, the present model has two input parameters; $\alpha$ is the introduction rate of new diseases on the network while the disease time $\tau$ corresponds to the duration any node is infectious with a disease.

A key element is that each node can only transmit one of its diseases at any given time step. Thus, a disease is less likely to spread from a node that carries many other diseases. The model can be run online as a java-applet at \textit{cmol.nbi.dk/models/disease/MultipleDiseases.html}.

\begin{figure}[htbp]
	\begin{center}
			\includegraphics[width=\columnwidth]{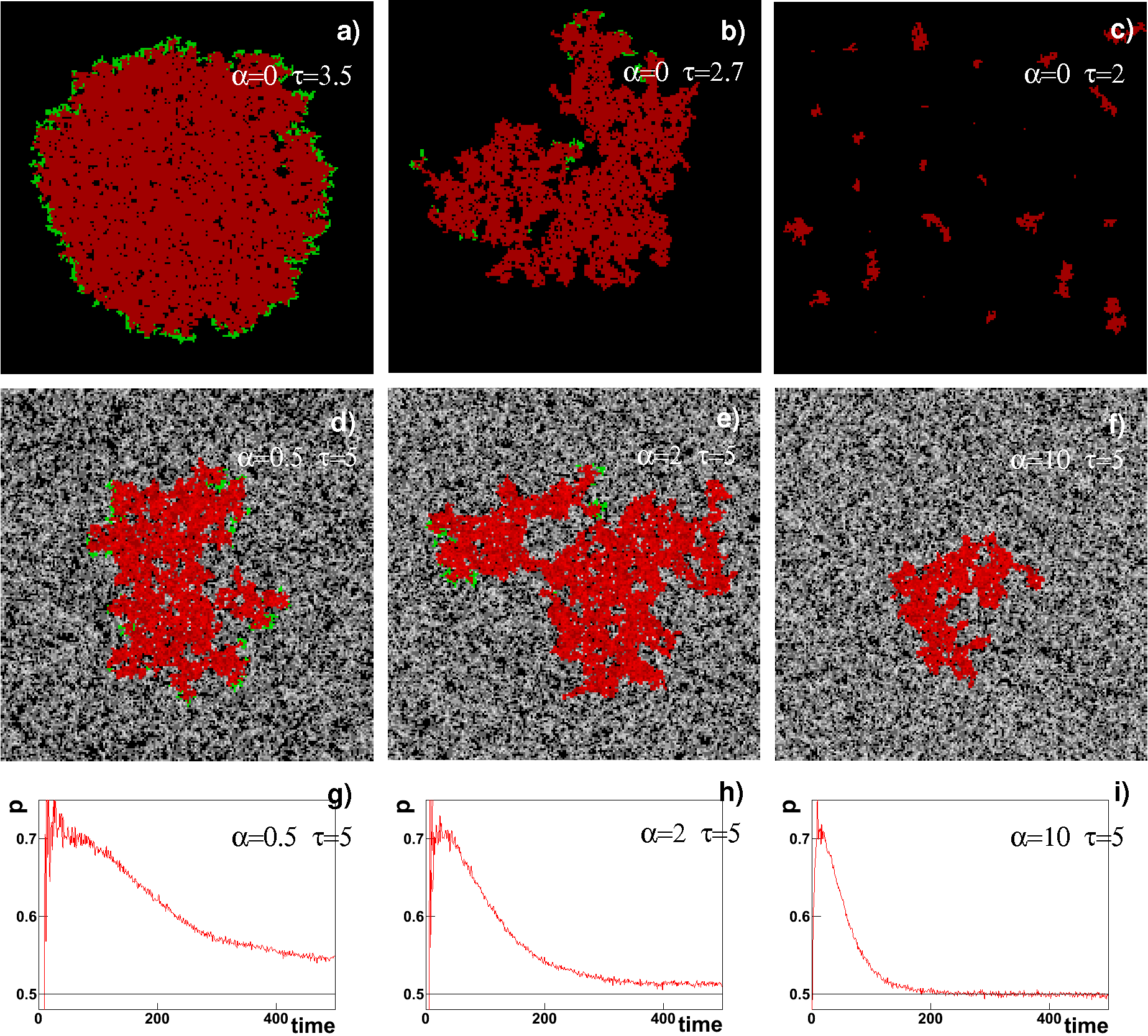}
		\caption{Spread of diseases on a lattice of size $L=256$ for different input parameters. Black nodes are healthy, and bright nodes carry many diseases. Accentuated clusters represent a particular disease. In $(a-c)$ only one disease is present $(\alpha \approx 0)$. If $\tau \approx 2.77$, diseases will spread in fractal shapes of critical percolation clusters. In the supercritical $(\tau >2.77)$ and subcritical $(\tau <2.77)$ cases, diseases will grow to span the entire network or quickly die out, respectively. In $(d-f)$, $\alpha>0$ the number of diseases per node will self-organize to a value where the transmitting probability is close to the critical threshold of percolation $p_c = \frac 12$. Thus, disease clusters have fractal shapes for a wide range of input parameters. The development of $p$ is shown in $(g-i)$.}
		\label{fig:2}
	\end{center}
\end{figure}

If a node is constantly infected with $k$ diseases, the probability that it tries to infect a given neighbor with a given disease before it is cured, can be found to be
\begin{eqnarray} \label{eqn:pConstantk}
	p = 1 - \exp \left(-\frac{\tau}{4k} \right).
\end{eqnarray}
This probability corresponds to the percolation probability of a bond percolation system. When $\alpha \approx 0$, diseases are rare and infected nodes will have only one disease. When $\tau$ is large, this disease will have plenty of time to infect its neighbors, and will therefore spread in a circular manner with a broad rim of infected nodes and a solid interior of recovered nodes, very similar to the well-studied Eden growth \cite{Eden, Mollison, Martin}. When $\tau$ is small, this disease will rarely manage to infect a neighbor before dying out. For $\tau=4 \ln(2) \approx 2.77$, we see from~\eqref{eqn:pConstantk} that the probability of infecting a given neighbor is exactly $p_c = \frac 12$, which is the critical threshold of bond percolation on a 2d-lattice \cite{Sykes}. In this case, the spread of the disease becomes fractal-like and the regions of infected nodes are only one node thick. This behavior, which is shown in figure \ref{fig:2}(a)-\ref{fig:2}(c) for a system of size $L=256$, is well-studied for single disease models \cite{Grassberger}.

When $\alpha>0$ the analysis is made complicated by spatial and temporal variations in $k$ and, therefore, $p$. For $\tau>2.77$, the system will initially be supercritical, and all diseases will grow rapidly. Consequently, the average number of diseases per node $\left< k \right>$ will increase and the average probability $\left< p \right>$ over the lattice will decrease. If $\left< p \right>$ becomes less than $\frac 12$, most new diseases will only spread to a handful of nodes before dying out. Thus, $\left< k \right>$ will decrease and $\left< p \right>$ will increase. This negative feedback mechanism will drive the system to a state, with $\left< p \right>$ close to the critical percolation threshold, where disease sizes of all orders of magnitudes occur. In this state, the number of diseases per node is close to Poisson distributed across the lattice, but with both spatial and temporal correlations in $k$. A high $\tau$ will result in many diseases per node and a high $\alpha$ will make the system self-organize faster, but for a wide range of both parameters diseases will spread in fractal-like shapes, as can be seen in figure \ref{fig:2}(d)-\ref{fig:2}(f).

The average probability $\left< p \right>$ of transmitting a disease can be measured directly by monitoring how many neighbors each node on average tries to infect with a disease, before it is cured. In figure \ref{fig:2}(g)-\ref{fig:2}(i) the development $\left< p \right>$ is shown for various input parameters. It is seen that $\left< p \right>$ converges to a value close to $\left< p \right> = \frac 12$.

\section{Critical exponents}
To compare the model to a percolation system, the clusters of recovered and immune nodes were investigated for different sets of input parameters. For each disease, its cluster diameter, mass and exterior perimeter were measured. 

\begin{figure}[tbp]
	\begin{center}
			\subfigure[Cluster mass scales with the diameter with exponent close to the fractal dimension of critical percolation $D \approx 1.896$.]{\label{fig:dim}\includegraphics[width=0.95\columnwidth]{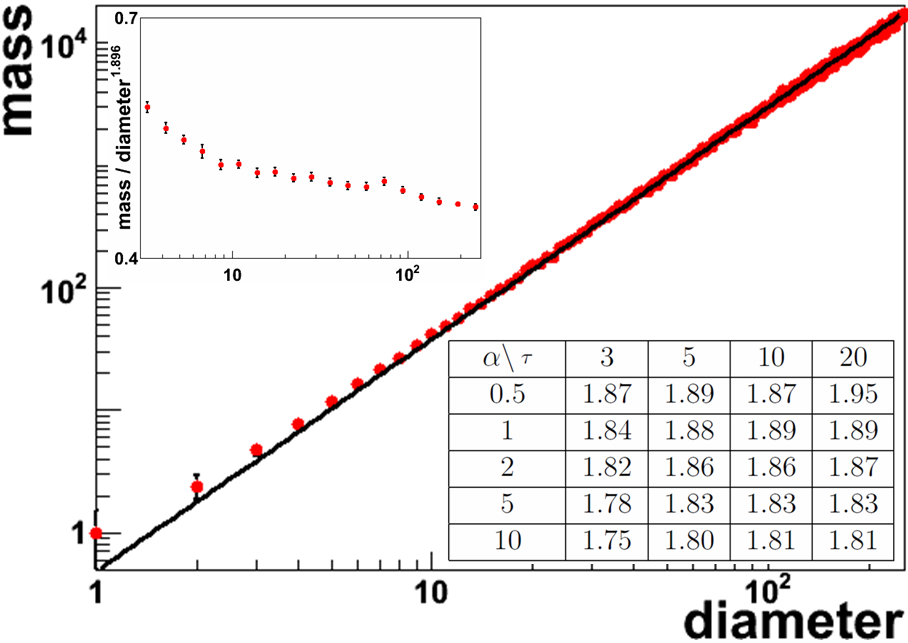}} \\
			\subfigure[External perimeter scales with the diameter with exponent close to that of critical percolation $D_e = \frac 43$.]{\label{fig:per}\includegraphics[width=0.95\columnwidth]{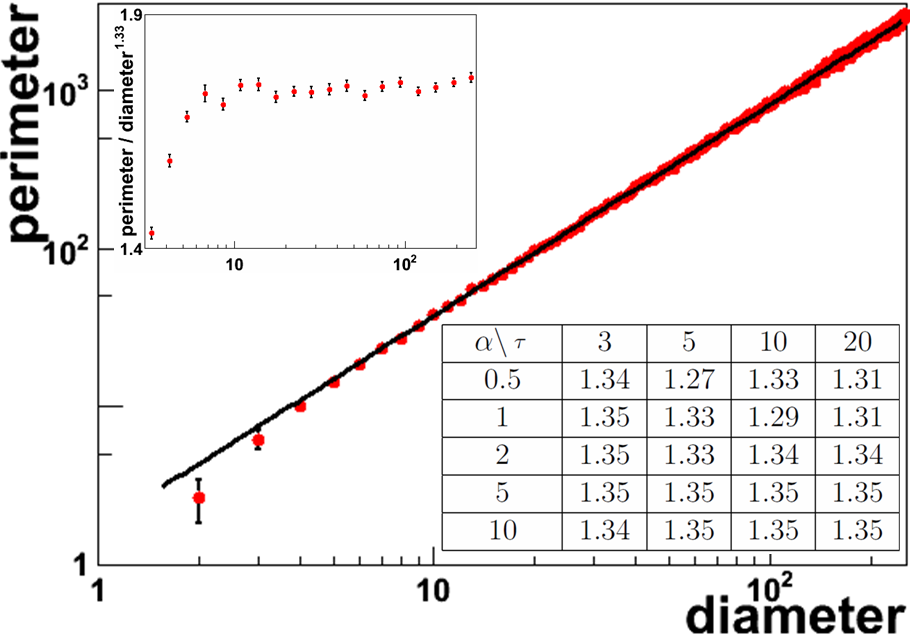}} \\
			\subfigure[Cluster size distribution falls off with exponent broadly distributed around that of critical percolation $ -1.05$.]{\label{fig:count}\includegraphics[width=0.95\columnwidth]{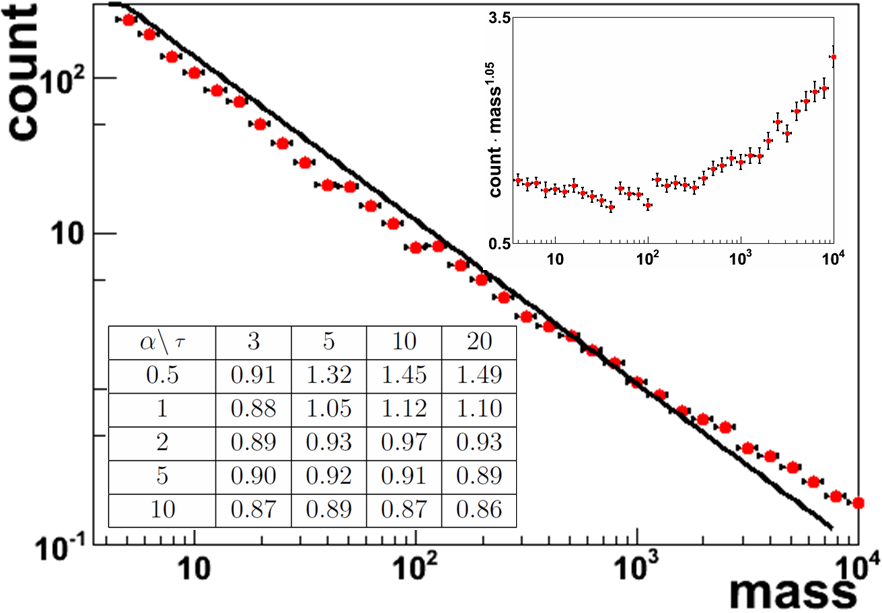}}
		\caption{Critical exponents for disease clusters when $\alpha = 2$ and $\tau = 5$. Black lines show the exponents for critical percolation, while the fitted values for the exponents are shown in the tables for different input parameters. The inserts show the experimental data normalized with respect to the critical percolation exponents.}
	\end{center}
	\label{fig:criticalExponents}
\end{figure}

For critical percolation the cluster mass scales with the diameter giving a fractal dimension of $D=\frac{91}{48} \approx 1.896$ \cite{Stauffer}. In figure \ref{fig:dim} it is seen that disease clusters have dimensions very close to this, for a wide range of input parameters. The fractal dimension is larger than $\frac{91}{48}$ when $\frac{\alpha}{\tau}$ is small and vice versa.  

The exterior perimeter is defined as the number of sites in the cluster that have one or more neighbors strictly outside the cluster. For critical percolation, it scales with the diameter with the critical exponent $D_e = \frac 43$ \cite{Stauffer}. In figure \ref{fig:per} this is seen to be very close to the scaling of disease clusters.

At the critical point, the cluster size distribution is expected to fall off with the critical exponent $-\frac{96}{91} \approx -1.05$, such that small diseases occur more frequently than large. In figure \ref{fig:count} it is seen that the disease clusters fall off with exponents broadly distributed around this value, with steeper exponents when $\frac{\alpha}{\tau}$ is small. Here, the chance of getting a disease spanning the entire network is large, but the chance of a large disease suddenly dying out is low.

\section{Discussion}
\begin{figure}[tbp]
	\centering
		\includegraphics[width=\columnwidth]{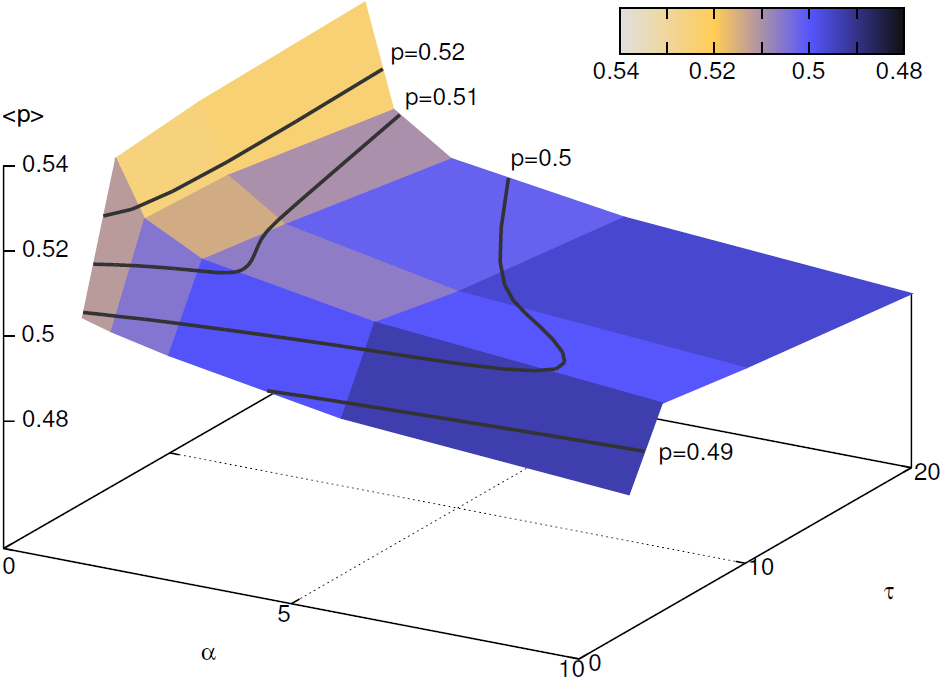}
	\caption{Steady state average transmitting probability $\left< p \right>$ shown for different input parameters. When $\tau>2.77$, the number of diseases per node self-organizes to a value, such that $\left< p \right>$ is close to the critical probability of percolation. For low $\alpha$, the system is slightly supercritical and, conversely, when $\alpha$ is high, the system is slightly subcritical. Fine-tuning of a parameter is necessary in order for the system to be truly critical. Note that the critical threshold is not necessarily at $p_c = \frac 12$ due to correlations in the number of diseases per node.}
	\label{fig:quasi}
\end{figure}

In figure \ref{fig:quasi}, the steady state probability $\left< p \right>$ is shown as a function of the input parameters. The observations agree well with the characteristics of quasi-criticality \cite{Bonachela}. The system self-organizes to a near-critical state, but a fine-tuning of a parameter (e.g. $\alpha$) is necessary in order for the system to be truly critical. When $\alpha$ is too low, diseases are transmitted with a probability somewhat larger that $p_c$, and the system is supercritical - the disease clusters become 'heavy' with a fractal dimension above $D_e=\frac{91}{48}$, an external perimeter dimension below $D=\frac{4}{3}$ and high probability of forming an epidemic. When $\alpha$ is too high, the system is subcritical with low $\left< p \right>$ and $D$, high $D_e$, and with low probability of forming an epidemic.

It should be emphasized that, due to correlations in the number of diseases per node, the critical threshold of the multiple disease model is not necessarily equal to that of critical percolation $p_c = \frac 12$. Diseases will tend to 'get stuck' and accumulate in regions where there are already many diseases, while they will quickly 'sweep over' regions with few diseases. 

In the model, a node carrying $k$ diseases will have a probability of $\frac 1k$ to pass on each of them. This coupling mechanism between diseases is not based on empirical evidence, but merely reflects that diseases compete against each other. However, the self-organization is robust to changes in the dynamics as long as diseases inhibit each other. For instance, if a node carrying $k$ diseases has a probability of $\frac 1{k^2}$ to pass on each of them, the case $\alpha = 1$ and $\tau = 5$ gives clusters with $D=1.875$, $D_e=1.331$, which indeed is close to the critical exponents of percolation.

\section{Conlusion}
Using a recently developed framework for spread of many diseases \cite{Sneppen} we have presented a simple multiple disease model that exhibits self-organized quasi-critical percolation \cite{Bonachela}. The model is based only on local information, having no global control mechanism or separation of time scale. Furthermore, the main feature of the self-organization is robust to changes in the extent to which multiple diseases inhibit each other. The basic mechanism employed in this model may be applicable to other systems, where ``new" has an intrinsic advantage over ``old", and where transmission capacity is limited. 
Thus the model may be equally valid as a rumor spreading model, where rumors compete for attention and become locally outdated. In this framework the predicted self-organization may help to explain the broad distributions found in human social activities. 


%

\end{document}